\RequirePackage[mathlines]{lineno}
\documentclass[amsmath,amssymb,twocolumn,prd,floatfix,showpacs, nofootinbib]{revtex4-1}
\usepackage{tabularx}  
\usepackage{bm, color} 
\usepackage{overpic,subfigure} 
\usepackage{multirow}
\usepackage{array}
\usepackage{dcolumn} 
\usepackage[symbol]{footmisc}
\usepackage{epstopdf}
\usepackage{lineno}
\usepackage{hyperref}
\usepackage{amsmath}
\usepackage{float}

\newcommand{\BESIIIorcid}[1]{\href{https://orcid.org/#1}{\hspace*{0.1em}\raisebox{-0.45ex}{\includegraphics[width=1em]{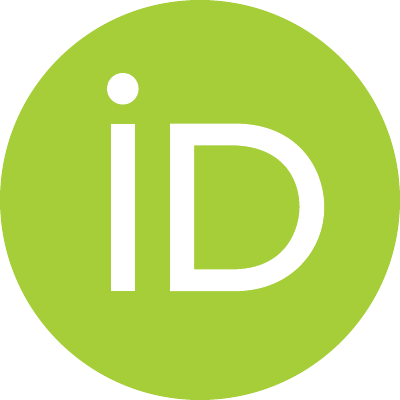}}}}

\hypersetup{hypertex=true,
	        colorlinks = true,
            linkcolor = blue,
            urlcolor = blue,
            citecolor = blue}

\begin{document}

\title{\boldmath Measurements of the branching fractions of $\chi_{cJ}\to 2K^+ 2K^- \omega$ and $\phi K^+ K^- \omega$ decays }

\author{
M.~Ablikim$^{1}$\BESIIIorcid{0000-0002-3935-619X},
M.~N.~Achasov$^{4,c}$\BESIIIorcid{0000-0002-9400-8622},
P.~Adlarson$^{81}$\BESIIIorcid{0000-0001-6280-3851},
X.~C.~Ai$^{86}$\BESIIIorcid{0000-0003-3856-2415},
R.~Aliberti$^{39}$\BESIIIorcid{0000-0003-3500-4012},
A.~Amoroso$^{80A,80C}$\BESIIIorcid{0000-0002-3095-8610},
Q.~An$^{77,64,\dagger}$,
Y.~Bai$^{62}$\BESIIIorcid{0000-0001-6593-5665},
O.~Bakina$^{40}$\BESIIIorcid{0009-0005-0719-7461},
Y.~Ban$^{50,h}$\BESIIIorcid{0000-0002-1912-0374}
% ...
\\
\vspace{0.2cm}
(BESIII Collaboration)\\
\vspace{0.2cm} {\it
$^{1}$ Institute of High Energy Physics, Beijing 100049, People's Republic of China\\
% ...
$^{4}$ Budker Institute of Nuclear Physics SB RAS (BINP), Novosibirsk 630090, Russia\\
% ...
$^{39}$ Johannes Gutenberg University of Mainz, Johann-Joachim-Becher-Weg 45, D-55099 Mainz, Germany\\
$^{40}$ Joint Institute for Nuclear Research, 141980 Dubna, Moscow region, Russia\\
% ...
$^{50}$ Peking University, Beijing 100871, People's Republic of China\\
% ...
$^{62}$ Southeast University, Nanjing 211100, People's Republic of China\\
% ...
$^{64}$ State Key Laboratory of Particle Detection and Electronics, Beijing 100049, Hefei 230026, People's Republic of China\\
% ...
$^{77}$ University of Science and Technology of China, Hefei 230026, People's Republic of China\\
% ...
$^{80}$ University of Turin and INFN, (A)University of Turin, I-10125, Turin, Italy; (B)University of Eastern Piedmont, I-15121, Alessandria, Italy; (C)INFN, I-10125, Turin, Italy\\
$^{81}$ Uppsala University, Box 516, SE-75120 Uppsala, Sweden\\
% ...
$^{86}$ Zhengzhou University, Zhengzhou 450001, People's Republic of China
\\
\vspace{0.2cm}
$^{\dagger}$ Deceased\\
% ...
$^{c}$ Also at the Novosibirsk State University, Novosibirsk, 630090, Russia\\
% ...
$^{h}$ Also at State Key Laboratory of Nuclear Physics and Technology, Peking University, Beijing 100871, People's Republic of China\\
% ...
}
}		

\begin{abstract}
Using a data sample of $(2712.4 \pm 14.3) \times 10^{6}$ $\psi(3686)$ events collected with the BESIII detector operating at the BEPCII collider, we report the first observation of the decays $\chi_{cJ}\to 2K^+ 2K^- \omega$ and $\chi_{cJ}\to \phi K^{+}K^{-} \omega$ ($J = 0,1,2$) via the radiative transitions $\psi(3686) \to \gamma \chi_{cJ}$. The branching fractions of these decays are measured for the first time, and the statistical significance for each signal exceeds $10\sigma$.
\end{abstract}

\maketitle

\section{Introduction}
Experimental studies of charmonium states and their decay properties play an important role in testing quantum chromodynamics (QCD) models and QCD based calculations~\cite{ref::work1, ref::work2, ref::work4}. In the quark model, the $\chi_{cJ}~(J = 0,1,2)$ mesons are the $^{3}P_{J}$ charmonium states. Experimentally, they have not been investigated as extensively as the vector charmonium states $J/\psi$ and $\psi(3686)$. At the BESIII experiment, $\chi_{cJ}$ mesons are rarely produced directly in $e^{+}e^{-}$ collisions; however, they are abundantly produced through the radiative decays of $\psi(3686)$, whose branching fractions are approximately $9\%$ for each $\chi_{cJ}$ state~\cite{ref::pdg2025}, thereby providing a large data sample for study.
At present, the sum of all measured branching fractions of $\chi_{cJ}$ decays is still far below unity, and knowledge of exclusive multi-body decays remains limited. Therefore, the search for additional multi-body $\chi_{cJ}$ decay modes is essential for improving our understanding of their decay dynamics and for identifying possible intermediate resonant structures.

In 2009, 2012, and 2021, BESIII collected a total of $(2712.4 \pm 14.3)\times10^{6}$ $\psi(3686)$ events~\cite{ref::work3}. 
Owing to the abundant production of $\chi_{cJ}$ states in the radiative decays of $\psi(3686)$, the BESIII detector provides an ideal environment for searching for new multi-body $\chi_{cJ}$ decay modes.
By analyzing this data sample, we report the first observations and measurements of branching fraction of the decays $\chi_{cJ}\to 2K^+ 2K^- \omega$ and $\chi_{cJ}\to \phi K^{+} K^{-} \omega$.

\section{BESIII DETECTOR AND MONTE CARLO SIMULATION}
\label{sec:BES}
The BESIII detector~\cite{ref::detector} records symmetric $e^+e^-$ collisions provided by the BEPCII storage ring~\cite{ref::collider}
in the center-of-mass energy range from 1.84 to 4.95~GeV, with a peak luminosity of $1.1 \times 10^{33}\;\text{cm}^{-2}\text{s}^{-1}$ achieved at $\sqrt{s} = 3.773\;\text{GeV}$.

The cylindrical core of the BESIII detector covers 93\% of the full solid angle and consists of a helium-based multilayer drift chamber~(MDC),
a plastic scintillator time-of-flight system~(TOF), and a CsI (Tl) electromagnetic calorimeter~(EMC).
These components are enclosed in a superconducting solenoidal magnet that provides a 1.0~T magnetic field. The solenoid is supported by an octagonal flux-return yoke with resistive plate counter muon
identification modules interleaved with steel. The charged-particle momentum resolution at $1~{\rm GeV}/c$ is $0.5\%$, and the ${\rm d}E/{\rm d}x$
resolution is $6\%$ for electrons from Bhabha scattering. The EMC measures photon energies with a resolution of $2.5\%$ ($5\%$) at $1$~GeV in the barrel (end-cap) region. The time resolution in the TOF barrel region is 68~ps, while that in the end-cap region was 110~ps.
The end-cap TOF system was upgraded in 2015 using multi-gap resistive plate chamber technology, providing a time resolution of 60~ps~\cite{Tof1,Tof2,Tof3}.
About $83\%$ of the data used in this analysis benefit from this upgrade.

Simulated data samples produced with a {\sc
geant4}-based~\cite{Geant4} Monte Carlo (MC) package, which
includes the geometric description~\cite{Geometric} of the BESIII detector and the
detector response, are used to determine detection efficiencies
and to estimate backgrounds. The simulation models the beam
energy spread and initial state radiation (ISR) in the $e^+e^-$
annihilations with the generator {\sc
kkmc}~\cite{Jadach01, Jadach02}.
The inclusive MC sample includes the production of the
$\psi(3686)$ resonance, the ISR production of the $J/\psi$, and
the continuum processes incorporated in {\sc
kkmc}~\cite{Jadach01}.
All particle decays are modelled with {\sc
evtgen}~\cite{Lange01,Lange02}, using branching fractions
either taken from the
Particle Data Group (PDG)~\cite{ref::pdg2025} when available,
or otherwise estimated with {\sc lundcharm}~\cite{Lundcharm00, Lundcharm01}.
Final state radiation
from charged final state particles is incorporated using the {\sc photos} package~\cite{PHOTOS}.
An inclusive MC sample of $2.7\times10^{9}$ generic $\psi(3686)$ events is used to study background.
A signal MC sample of 500,000 events for the decay $\chi_{cJ} \to 2K^+ 2K^- \omega$ is generated, including potential resonances according to the fractions listed in Table~\ref{mixmcfrac}.
The other independent signal MC sample containing 500,000 events for $\chi_{cJ} \to \phi K^{+}K^{-} \omega$ is generated according to phase space (PHSP).
The $\psi(3686) \to \gamma \chi_{cJ} ~(J = 0,1,2)$ decays are generated with the P2GCJ~\cite{Lange01} models.
The decays $\phi \to K^{+}K^{-}$ are generated by the VSS~\cite{VSS} model, which describes a vector meson decaying to a pair of pseudoscalar mesons.
The decays $\omega \to \pi^{+}\pi^{-}\pi^{0}$  are generated with the OMEGA\_DALITZ~\cite{Lange01} model.
All other signal decays are generated according to PHSP.

\begin{table}[htbp]
\caption{Fractions (in \%) of different sub-processes in $\chi_{cJ} \to 2K^+ 2K^- \omega$, where the uncertainties are statistical only. No other significant resonances are observed in data due to the limited sample size; the fractions of $2K^+ 2K^-$ (non-resonant) and $\phi K^+ K^-$ in the simulation were determined iteratively from the corresponding invariant-mass distributions as discussed in Sec.~\ref{bfs}.}
		\centering
        \scalebox{0.9}{
		\begin{tabular}{c|c|c|c}
			\hline \hline 
			Resonance source &$\chi_{c0}$      &$\chi_{c1}$   &$\chi_{c2}$            \\ \hline
			$\chi_{cJ} \to 2K^+ 2K^- \omega$ (non-resonant)      & $45.0 \pm 2.4$ & $44.7 \pm 3.5$ & $38.8 \pm 3.3$ \\ \hline
			$\chi_{cJ} \to \phi K^{+}K^{-} \omega$               & $55.0 \pm 2.4$ & $55.3 \pm 3.5$ & $61.2 \pm 3.3$ \\ \hline
			\hline 
		\end{tabular}
}
		\label{mixmcfrac}
	\end{table}

\section{EVENT SELECTION}
\label{sec:selection}

In this analysis, the $\phi$, $\omega$, and $\pi^{0}$ mesons are reconstructed via their decays $\phi \to K^+ K^-$, $\omega \to \pi^+ \pi^- \pi^0$, and $\pi^0 \to \gamma \gamma$, respectively. The branching fractions of $\chi_{cJ} \to 2K^+ 2K^- \omega$ and $\chi_{cJ} \to \phi K^{+}K^{-} \omega$ are measured via the radiative decays of $\psi(3686) \to \gamma \chi_{cJ}$. The signal candidates are required to have six charged tracks with zero net charge and at least three photon candidates.

Charged tracks in an event are required to have a polar angle $\theta$ with respect to the beam direction within the MDC acceptance $|\cos \theta|< 0.93$, and a distance of closest approach to the IP within 10\,cm along the beam direction and 1\,cm in the plane transverse to the beam direction. Particle identification~(PID) for charged tracks is based on the measurements of the d$E$/d$x$ in the MDC and the flight time in the TOF, which are combined to form a likelihood $\mathcal{L}(h)$~($h$ = $\pi$ and $K$) for each hadron $h$ hypothesis. Charged tracks with $\mathcal{L}(K)>\mathcal{L}(\pi)$ are identified as kaons, while those with $\mathcal{L}(\pi)>\mathcal{L}(K)$ are identified as pions.

Photon candidates are identified using showers in the EMC. The deposited energy of each shower must be greater than 25~MeV in the barrel region ($|\cos \theta|< 0.80$), where $\theta$ denotes the polar angle with respect to the MDC axis, and greater than 50~MeV in the end-cap region ($0.86 <|\cos \theta|< 0.92$). To exclude showers that originate from charged tracks, the angle between the EMC shower and the position of the closest charged track at the EMC must be greater than 10 degrees, as measured from the interaction point (IP). To suppress electronic noise and showers unrelated to the event, the difference between the EMC time and the event start time is required to be within [0, 700]\,ns.

In order to suppress backgrounds and improve the mass resolution, a five-constraint (5C) kinematic fit is performed under the hypothesis of $\psi(3686) \to \gamma 2K^+ 2K^- \pi^{+}\pi^{-}\pi^{0}$.
The fit constrains the total four-momentum of the final-state particles to that of the colliding beams, and an additional constraint requires the invariant mass of the photon pair to equal the nominal $\pi^{0}$ mass~\cite{ref::pdg2025}.
If more than one photon combination yields a convergent fit, the candidate with the smallest $\chi^{2}_{\rm 5C}$ is retained; events with $\chi^{2}_{\rm 5C}<40$ are kept for further analysis.

Potential background contributions from $\psi(3686)$ decays are studied by analyzing the inclusive MC sample with the generic event type analysis tool TopoAna~\cite{topo}. No significant peaking backgrounds containing both $\chi_{cJ}$ and $\omega$ mesons are observed. To suppress the background from the decay $\psi(3686) \to \pi^{+}\pi^{-}J/\psi$, a requirement is imposed on the recoil mass of $\pi^{+}\pi^{-}(RM(\pi^{+}\pi^{-}$)), defined as
\begin{equation}
RM(\pi^{+}\pi^{-}) = \sqrt{[\sqrt{s} - (E_{\pi^{+}} + E_{\pi^{-}})]^2 - (\vec{p}_{\pi^+} + \vec{p}_{\pi^-})^2},
\end{equation}
where $E_{\pi^{+}(\pi^-)}$ and $\vec{p}_{\pi^+(\pi^-)}$ are the energy and momentum of the positive (negative) pions, respectively. We require $|RM(\pi^{+}\pi^{-}) - M(J/\psi)| \textgreater 0.015$ GeV/$c^{2}$, where $M(J/\psi)$ is the nominal mass of $J/\psi$~\cite{ref::pdg2025}. 
Continuum background is evaluated with a data sample collected at $\sqrt s = 3.650$ GeV with an integrated luminosity of 454.33 pb$^{-1}$~\cite{lum}. Only a few events survive the selection, so this contribution is negligible.

\section{Branching fraction}
\label{bfs}
\subsection{$\chi_{cJ} \to 2K^+ 2K^- \omega$}
\label{4kfit}
To determine the signal yields of the decays $\chi_{cJ} \to 2K^+ 2K^- \omega$, a two-dimensional (2D) fit is performed to the distribution of $M(2K^+ 2K^- \omega)$ versus $M(\pi^{+}\pi^{-}\pi^{0})$ for the selected candidates in data. 
In the 2D fit, events are divided into four categories: the `signal' category contains candidates where both $2K^+ 2K^- \omega$ from the $\chi_{cJ}$ decay and $\pi^{+}\pi^{-}\pi^{0}$ from the  $\omega$ decay; `BKGI' contains candidates where $2K^+ 2K^- \omega$ from the $\chi_{cJ}$ decay while $\pi^{+}\pi^{-}\pi^{0}$ from continuum backgrounds; the `BKGII' contains candidates where $2K^+ 2K^- \omega$ from continuum backgrounds while $\pi^{+}\pi^{-}\pi^{0}$ from the $\omega$ decay; `BKGIII' contains candidates where both $2K^+ 2K^- \omega$ and $\pi^{+}\pi^{-}\pi^{0}$ from continuum backgrounds, as well as incompletely reconstructed events with similar final states. The corresponding probability-density functions (PDFs) are constructed as:
\begin{itemize}
\item signal: $S(x) \times S(y)$,
\item BKGI: $S(x) \times C(y)$,
\item BKGII: $C(x) \times S(y)$,
\item BKGIII: $C(x) \times C(y)$,
\end{itemize}
where $x$ and $y$ represent the two dimensions of the 2D fit, $S(i)$ are signal shapes derived from signal MC samples, $C(i)$ are second-order Chebyshev polynomials, and $i$ denotes the different dimension.

\begin{figure*}[htbp]
	\centering
	\includegraphics[width=15cm]{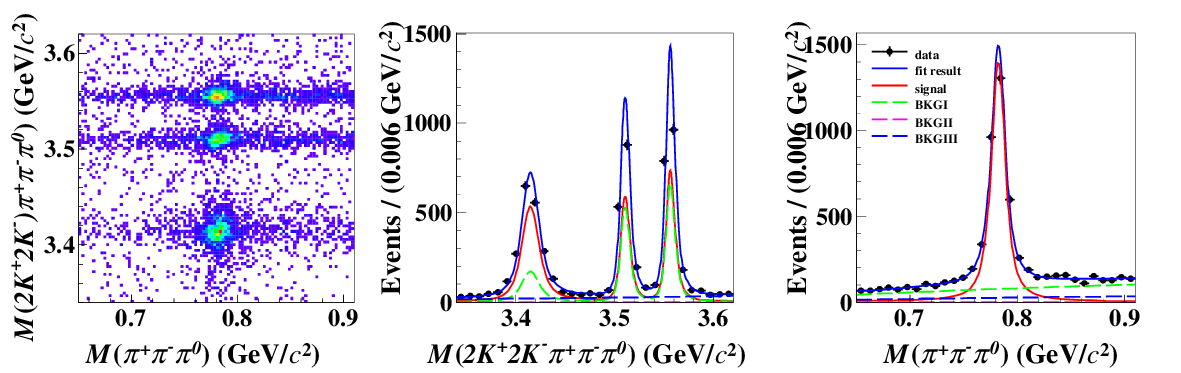}
	\caption{The first column is the distribution of $M(2K^+ 2K^-\pi^+\pi^-\pi^0)$ versus $M(\pi^+\pi^-\pi^0)$. The second and third columns show the projections of the 2D fit to the $M(2K^+ 2K^-\pi^{+}\pi^{-}\pi^{0})$ and $M(\pi^+\pi^-\pi^0)$ distributions, respectively. Data are shown as the dots with error bars, the blue solid curves are the fit results, the red solid curves are the fitted signal shapes, the green dashed curves are the BKGI events, the pink dashed curves are the BKGII events, and the blue dashed curves are the BKGIII events.}
	\label{fit:kkkk_sig}
\end{figure*}

Figure~\ref{fit:kkkk_sig} shows the 2D fit results for the signal candidates. The statistical significance of each of the $\chi_{c0,1,2} \to 2K^+ 2K^- \omega$, calculated using $\Delta(\ln\mathcal{L})=\ln\mathcal{L}_{\mathrm{max}}-\ln\mathcal{L}_{0}$ with $\Delta ndf ~=~ 1$, exceeds $10\sigma$. Here, $\ln\mathcal{L}_{\mathrm{max}}$ and $\ln\mathcal{L}_{0}$ are the maximum likelihoods obtained with and without the signal component in the fit, respectively, and $\Delta ndf$ is the change in the number of degrees of freedom.

\begin{figure*}[htbp]
\centering
		\includegraphics[width=0.255\textwidth]{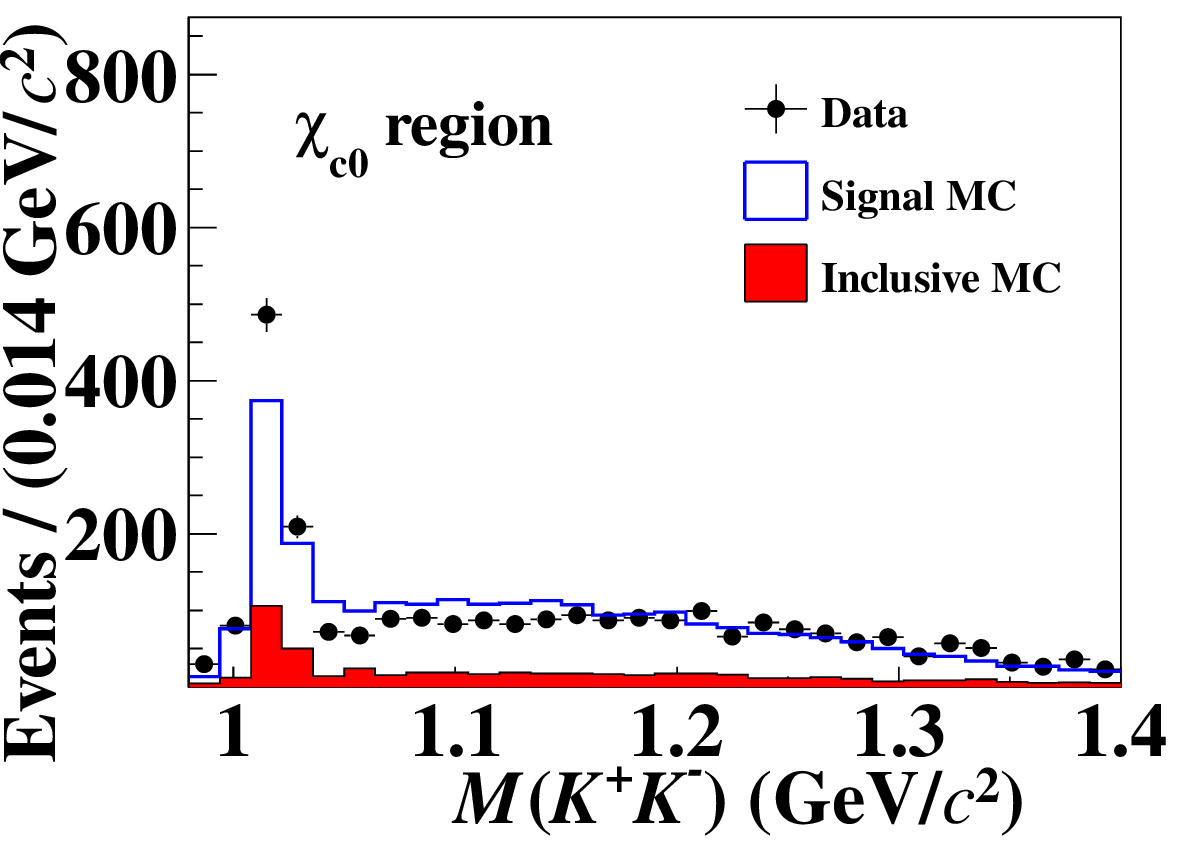}
		\includegraphics[width=0.255\textwidth]{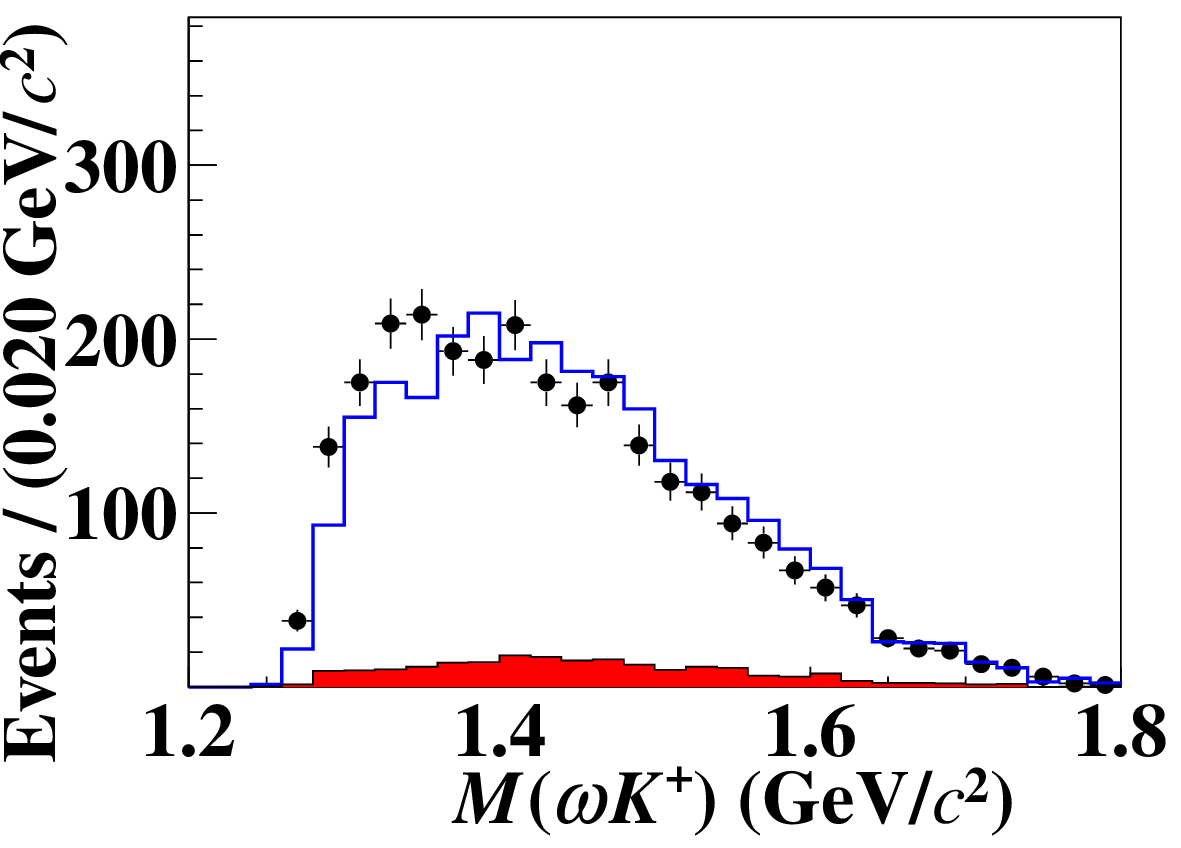}
		\includegraphics[width=0.255\textwidth]{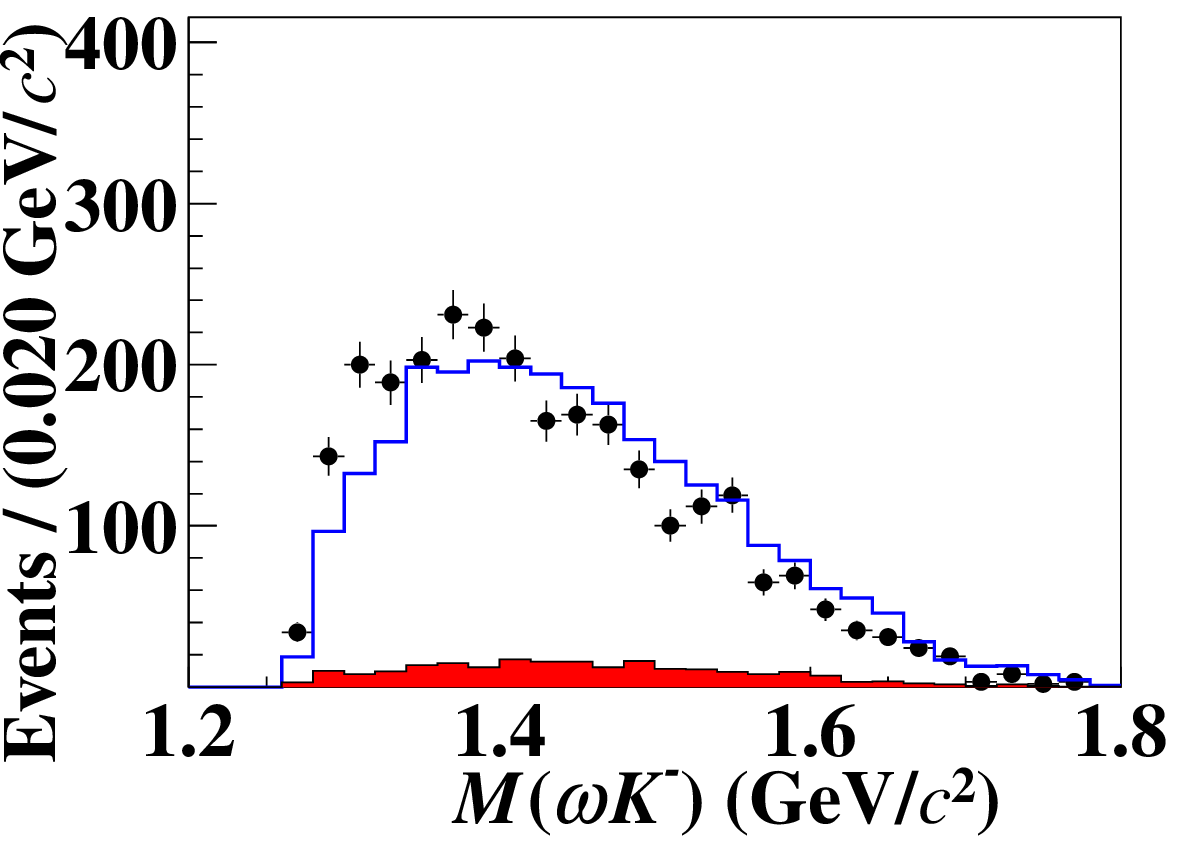}\\
		\includegraphics[width=0.255\textwidth]{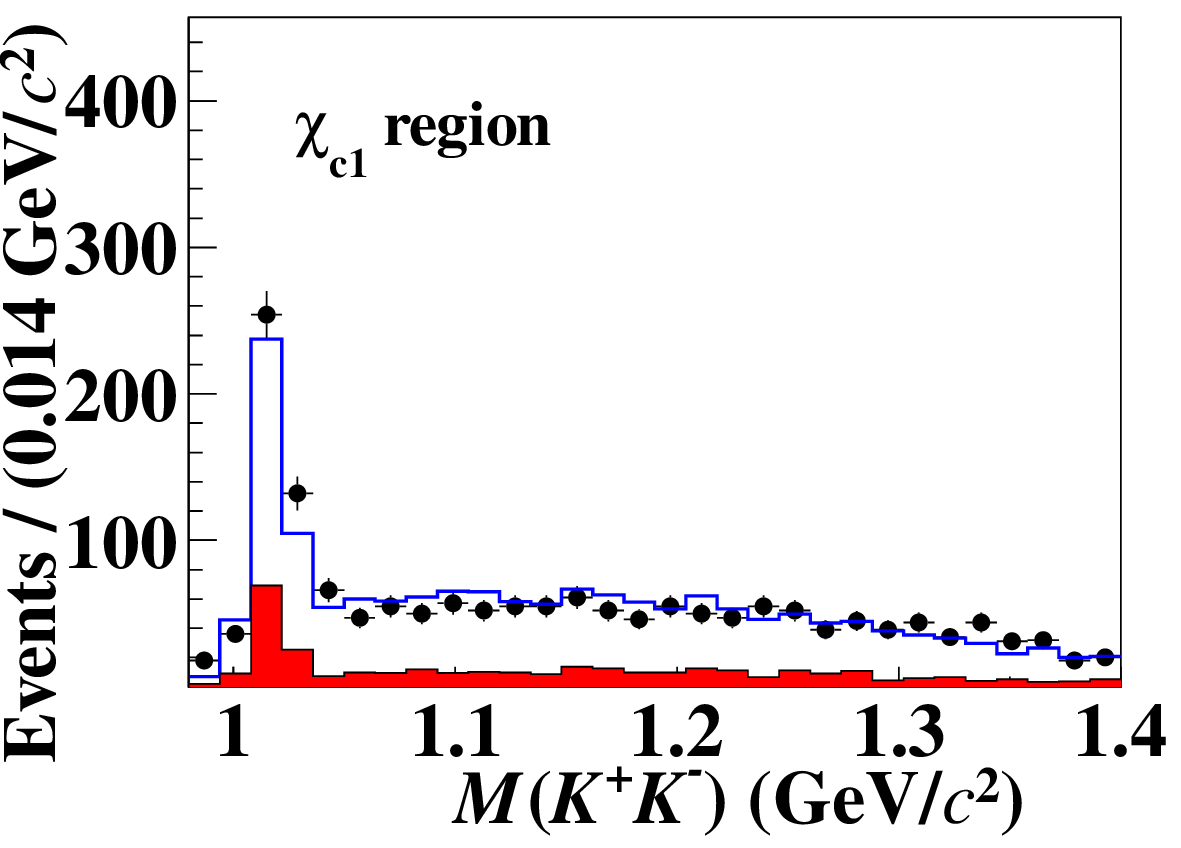}
		\includegraphics[width=0.255\textwidth]{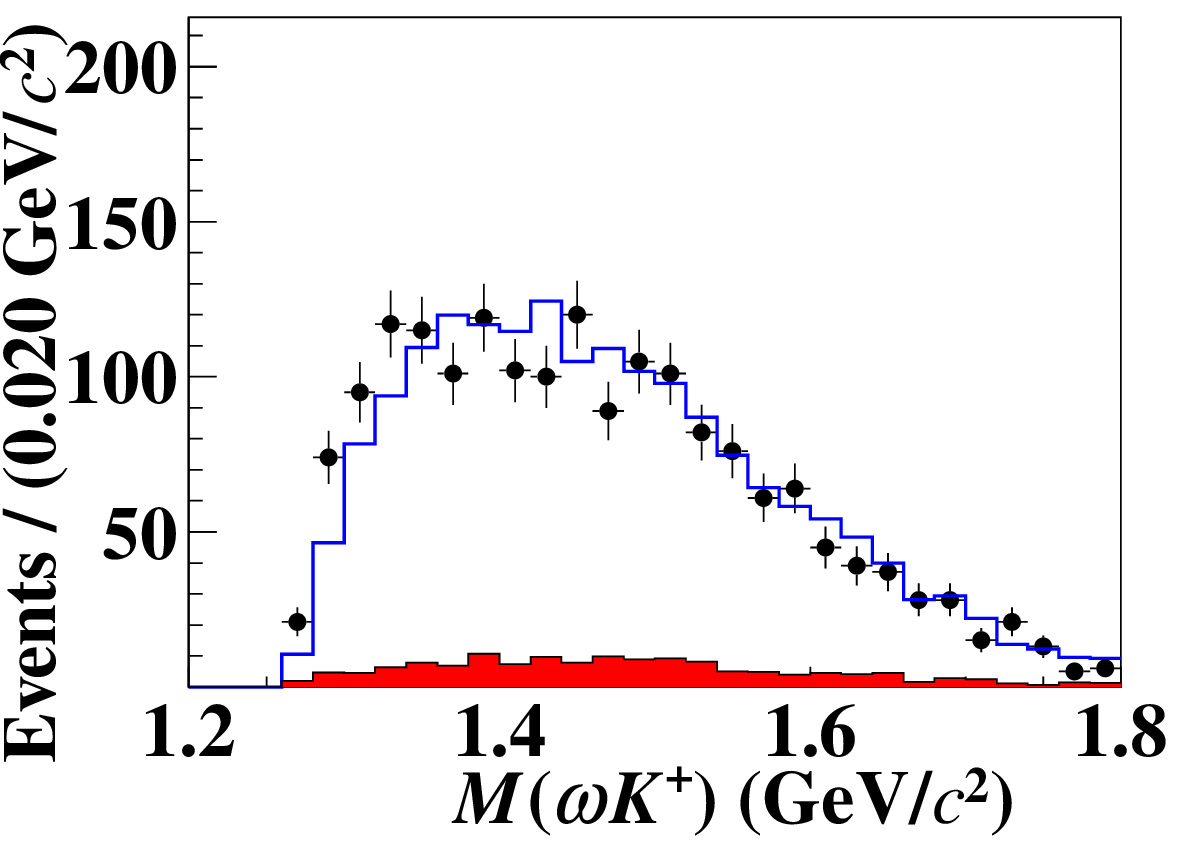}
		\includegraphics[width=0.255\textwidth]{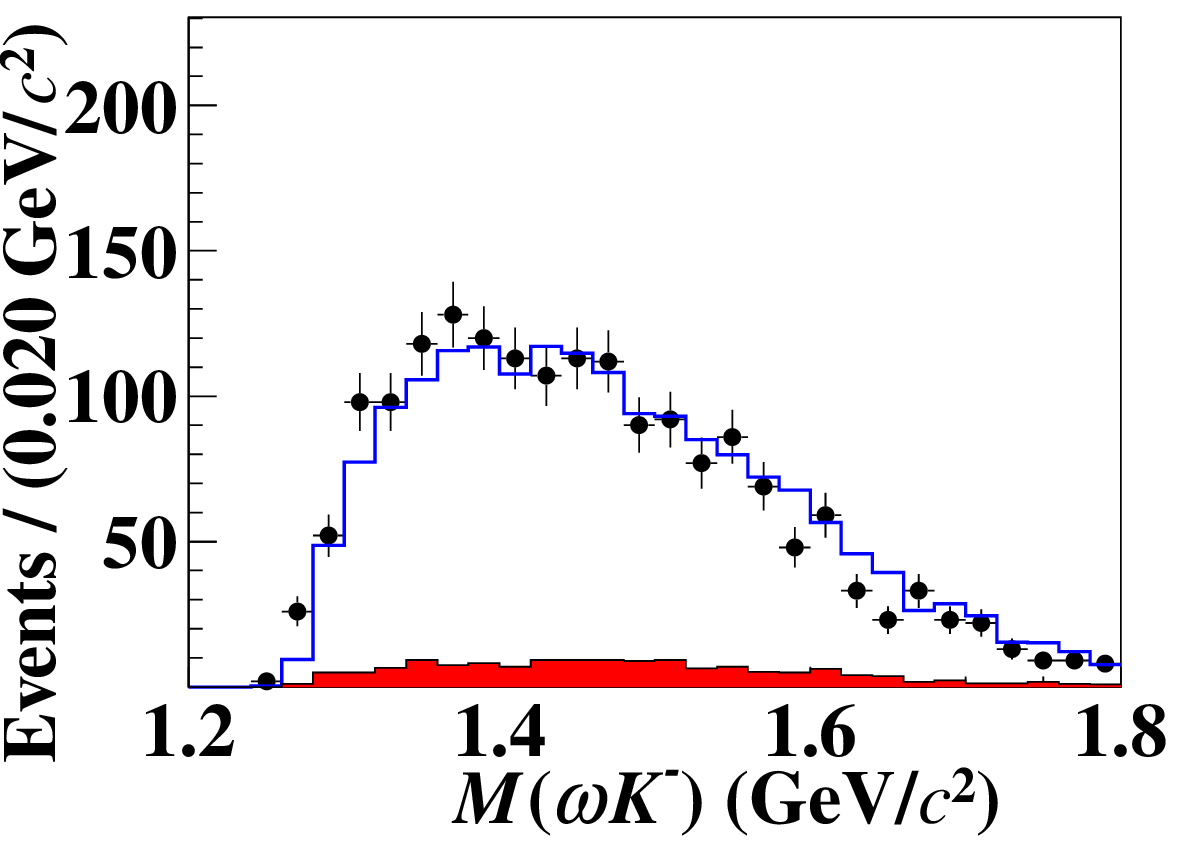}\\
		\includegraphics[width=0.255\textwidth]{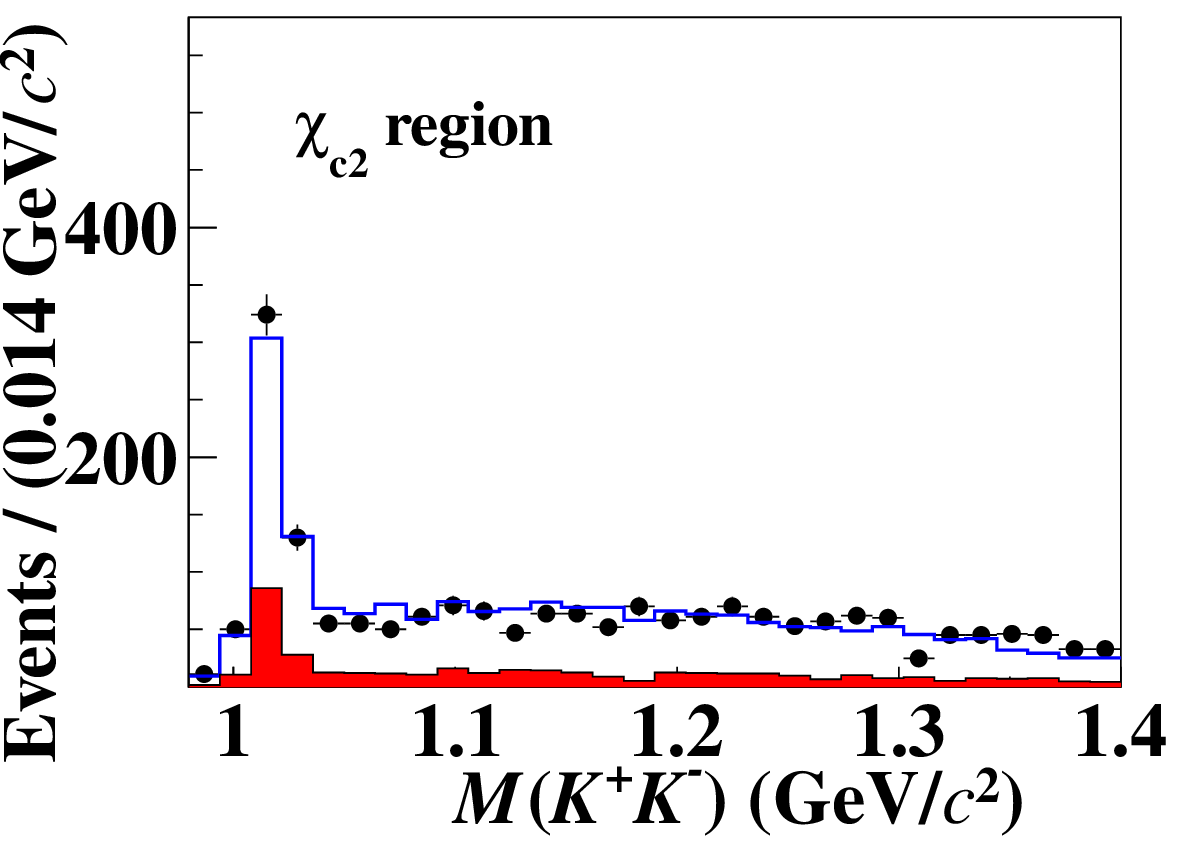}
		\includegraphics[width=0.255\textwidth]{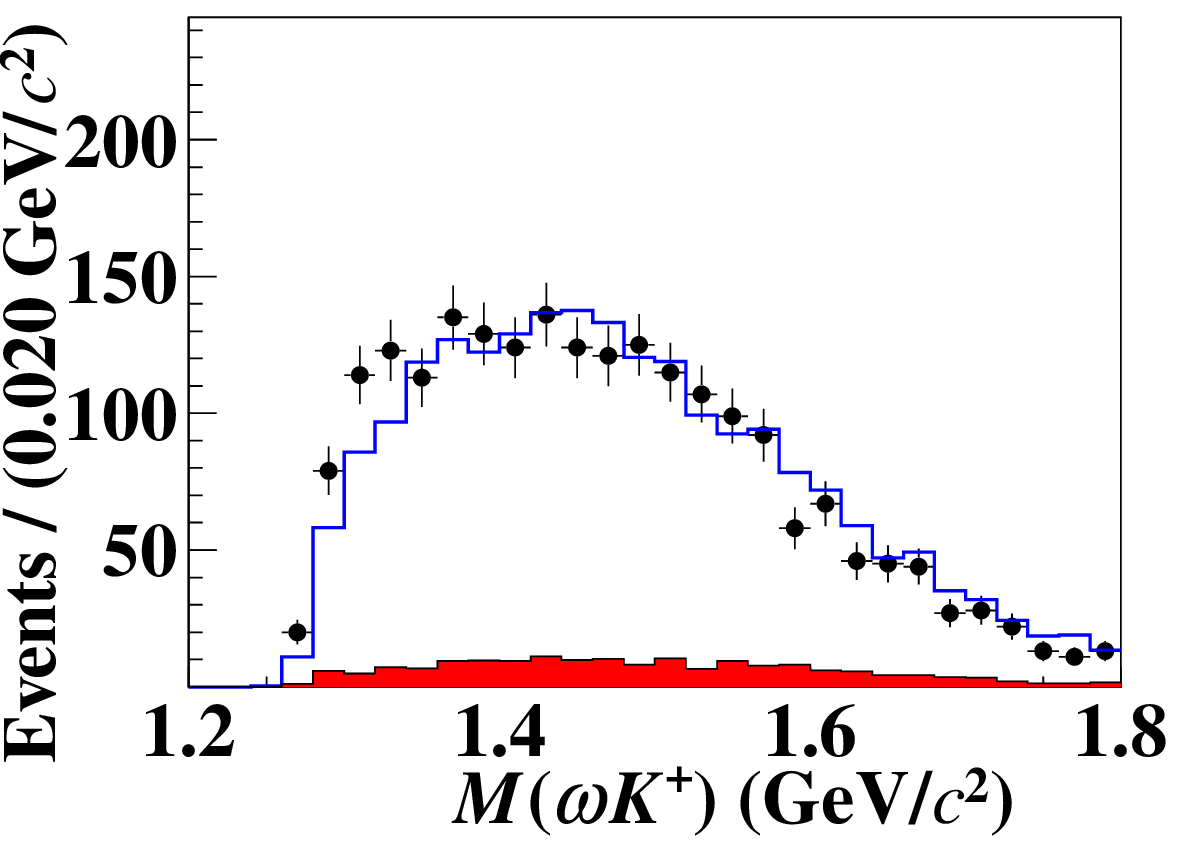}
		\includegraphics[width=0.255\textwidth]{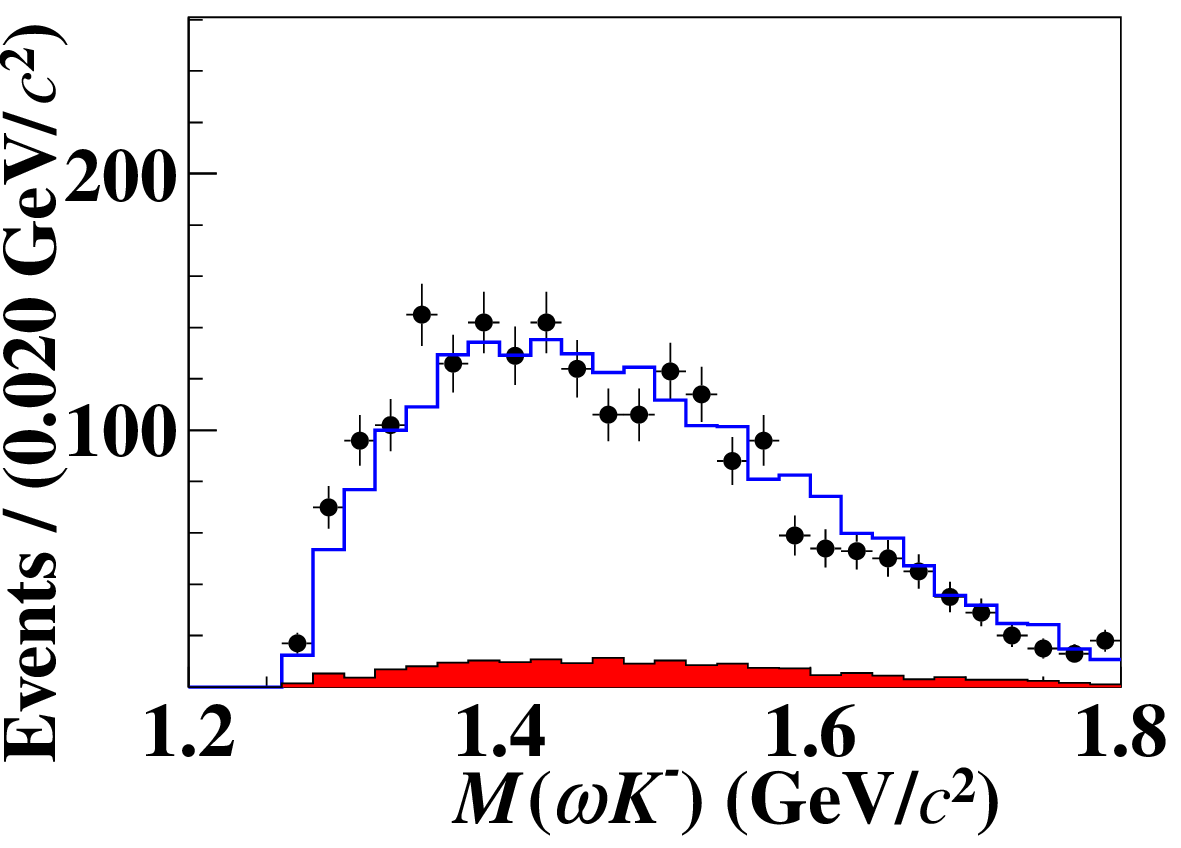}
        \caption{Comparisons of the invariant masses of possible two-body combinations of the selected candidates of $\chi_{cJ} \to 2K^+ 2K^- \omega$. The top, middle, and bottom rows are for events in the $\chi_{c0}$ mass region [3.385, 3.445] GeV/$c^{2}$, the $\chi_{c1}$ mass region [3.490, 3.530] GeV/$c^{2}$, and the $\chi_{c2}$ mass region [3.536, 3.576] GeV/$c^{2}$, respectively. The dots with error bars are data, the blue curves are the signal MC samples, and the red histograms are the inclusive MC samples. Events are required to be within the $\omega$ signal region, defined as $|M(\pi^{+}\pi^{-}\pi^{0}) - M(\omega)|~ \textless~ 0.025$ GeV/$c^{2}$.
        \label{comparision_kkkk:massesc0}
        }
\end{figure*}

\subsection{$\chi_{cJ} \to \phi K^{+}K^{-} \omega$}
\label{phikkfit}
Significant $\phi$ signals are observed in the $K^{+}K^{-}$ invariant-mass distributions (Fig.~\ref{comparision_kkkk:massesc0}), indicating the presence of $\chi_{cJ} \to \phi K^{+}K^{-} \omega$. This measurement constitutes an independent observation with respect to $\chi_{cJ} \to 2K^+ 2K^- \omega$ discussed above.
The $\phi$ signal region is defined as $M(K^{+}K^{-}) \in [1.005,\,1.035]\ \text{GeV}/c^{2}$, and the $\phi$ sideband region is defined as $M(K^{+}K^{-}) \in [1.045,\,1.075]\ \text{GeV}/c^{2}$.

\begin{figure}[htpb]
\centering
\includegraphics[width=8.5cm,height=8cm]{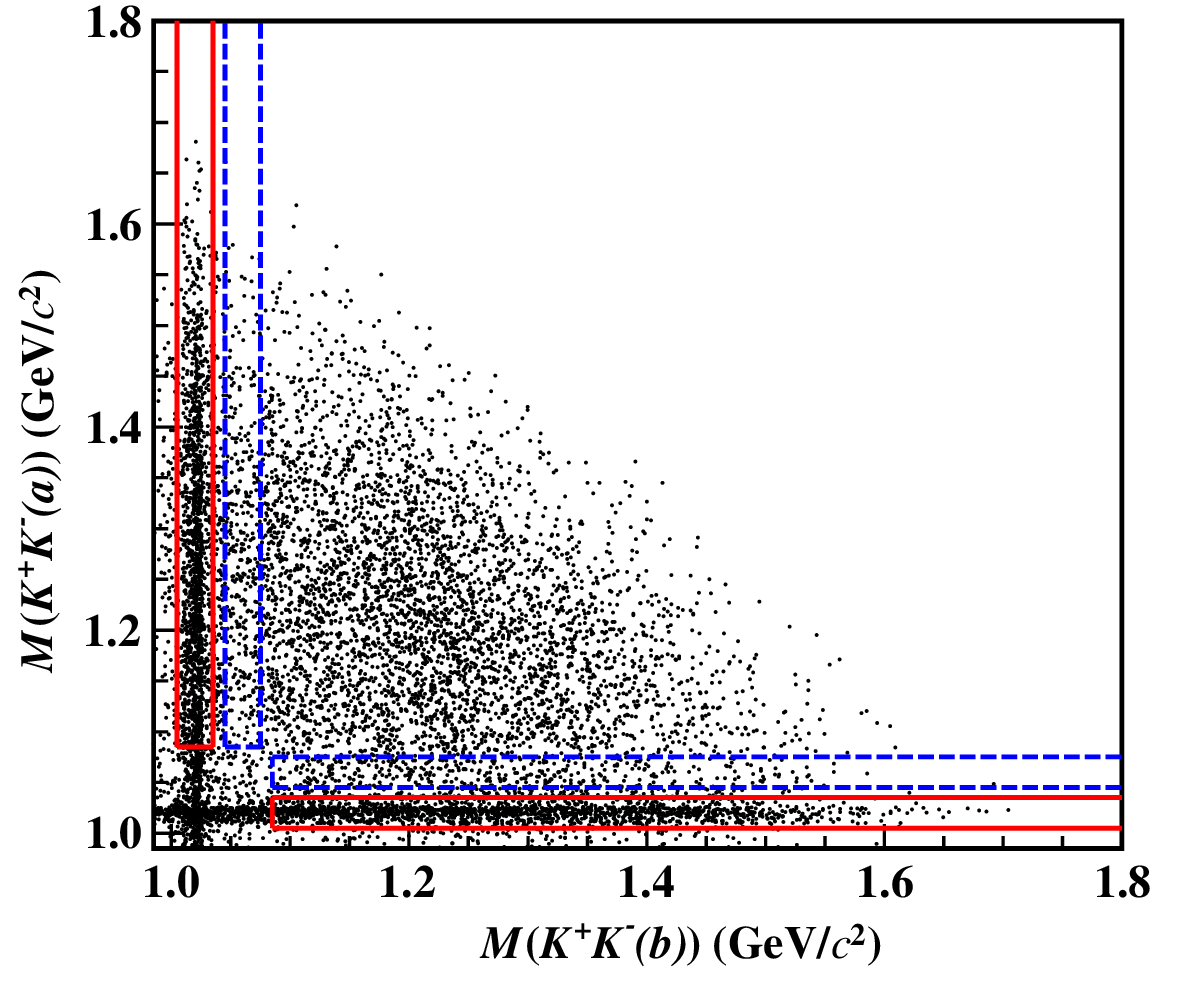}
\caption{The distribution of $M(K^{+}K^{-}(a))$ versus $M(K^{+}K^{-}(b))$ for the selected $\chi_{cJ} \to \phi K^{+}K^{-} \omega$ candidates in data, where the red boxes show the $\phi K^{+}K^{-}$ signal regions and the blue boxes show the $\phi K^{+}K^{-}$ sideband regions. Events have been required to be within the $\omega$ signal region, defined as $|M(\pi^{+}\pi^{-}\pi^{0}) - M(\omega)|~ \textless~ 0.025$ GeV/$c^{2}$.
			\label{plot:phi2k2D}
		}
\end{figure}

For the decays $\chi_{cJ} \to \phi K^{+}K^{-} \omega$ (with one $\phi$ and an additional $K^{+}K^{-}$ pair), all $K^+K^-$ combinations are retained.
We define the $M(K^{+}K^{-}(a))$ and $M(K^{+}K^{-}(b))$ denote the invariant masses of the randomly assigned $K^+K^-$ pairs.
A 2D analysis is performed by defining the 2D signal region as follows: one $K^{+}K^{-}$ pair lies within the $\phi$ signal region ($M(K^{+}K^{-}(a/b)) \in [1.005,\,1.035]\ \text{GeV}/c^{2}$), while the other $K^{+}K^{-}$ pair satisfies $M(K^{+}K^{-}(b/a)) > 1.085\ \text{GeV}/c^{2}$. The 2D sideband region is defined analogously, but with one $K^{+}K^{-}$ pair in the $\phi$ sideband region ($M(K^{+}K^{-}(a/b)) \in [1.045,\,1.075]\ \text{GeV}/c^{2}$) and the other $K^{+}K^{-}$ pair with $M(K^{+}K^{-}(b/a)) > 1.085\ \text{GeV}/c^{2}$. Figure~\ref{plot:phi2k2D} illustrates the 2D distribution of the $M(K^{+}K^{-})$ combinations for both signal and sideband regions, providing a visual distinction between correlated and combinatorial pairs. The peaking backgrounds in the $M(K^+K^-)$ distributions in Fig.~\ref{comparision_kkkk:massesc0} do not simultaneously produce peaking backgrounds in the $M(\pi^+\pi^-\pi^0)$ and $M(2K^+ 2K^- \omega)$ distributions. Other potential backgrounds are studied by analyzing the inclusive MC sample with the TopoAna~\cite{topo}. No significant peaking backgrounds with both $\chi_{cJ}$ and $\omega$ are observed.

\begin{figure*}[htbp]
	\centering
	\includegraphics[width=15cm]{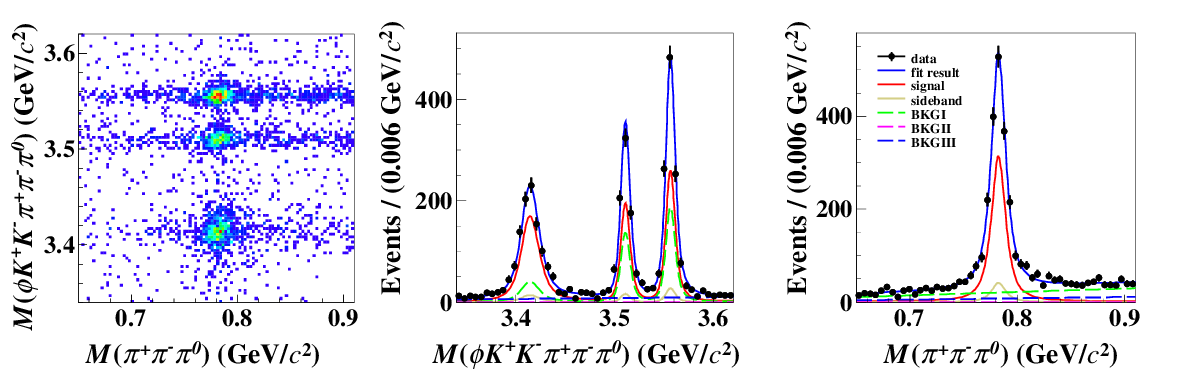}\\
	\includegraphics[width=15cm]{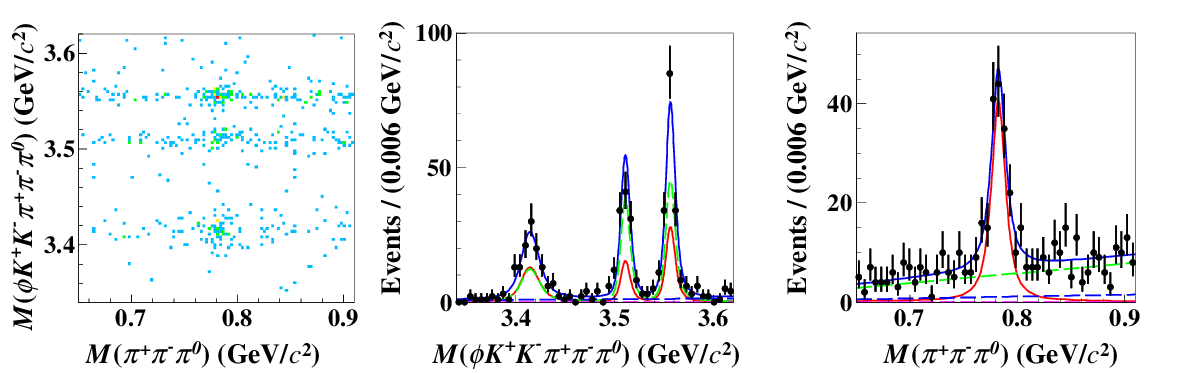}
	\caption{The first column is the distribution of $M(\phi K^{+}K^{-}\pi^+\pi^-\pi^0)$ versus $M(\pi^+\pi^-\pi^0)$. The second and third columns show the projections of the 2D simultaneous fit to the $M(\phi K^+K^-\pi^{+}\pi^{-}\pi^{0})$ and $M(\pi^+\pi^-\pi^0)$ distributions in the signal region (top row) and sideband region (bottom row). Data are shown as dots with error bars. For both the top and bottom figures, the blue solid curves are the total fit results, the red solid curves are the fitted signal shapes, the green dashed curves are the BKGI events, the pink dashed curves are the BKGII events, and the blue dashed curves are the BKGIII events. For the top figures, the brown solid curves are the normalized sideband contributions.}
	\label{fit:phikk_sigside}
\end{figure*}

The signal yields of $\chi_{cJ} \to \phi K^{+}K^{-} \omega$ are determined by a 2D simultaneous fit to the distribution of $M(\phi K^{+}K^{-} \pi^{+}\pi^{-}\pi^{0})$ versus $M(\pi^{+}\pi^{-}\pi^{0})$ for events in the $\phi$ signal and sideband regions of the selected candidates in data. 
In the simultaneous fit, the construction of the PDFs follows the approach described in Sec.~\ref{4kfit}.
The signal shapes of $\chi_{cJ}$ are described by the signal MC-simulated shapes, while the background shapes are described by a second-order Chebyshev polynomial.
Events in the $\phi$ sideband region are normalized using a scale factor $f$, which is evaluated from the ratios of the background yields in the $\phi$ signal and sideband regions and is found to be close to 1.0. Figure~\ref{fit:phikk_sigside} shows the results of simultaneous fit. 
The net signal yield $N_{\rm net}$ is determined as
\begin{equation}
N_{\rm net} = N_{\rm sig} - f \cdot N_{\rm sid}.
\end{equation}
From this fit, the signal yields of $\chi_{c0,1,2} \to \phi K^{+}K^{-} \omega$ are obtained, with the statistical significance larger than 10$\sigma$ for each $\chi_{cJ}$ decay.

A search for $\chi_{cJ} \to \phi \phi \omega$ was also performed; no significant signal is observed, consistent with C-parity violation, and results are not reported.

\subsection{Branching fractions}
Detection efficiencies for $\psi(3686) \to \gamma \chi_{cJ}$ with $\chi_{cJ} \to 2K^+ 2K^- \omega$ are determined using mixed signal MC samples with fractions of $\chi_{cJ} \to \phi K^{+}K^{-} \omega$, as summarized in Table~\ref{mixmcfrac}. The fractions of $\chi_{cJ} \to \phi K^{+}K^{-} \omega$ are determined by a 2D simultaneous fit shown in Fig.~\ref{fit:phikk_sigside}.

The branching fractions of $\chi_{cJ} \to X$ ($2K^+ 2K^- \omega$ and $\phi K^+K^- \omega$) are calculated using
\begin{equation}
\begin{aligned}
&\mathcal B(\psi(3686) \to \gamma \chi_{cJ}) \cdot \mathcal B(\chi_{cJ} \to X) =\frac{N_{\rm net}}{N_{\psi(3686)} \cdot \epsilon\cdot{\mathcal B(\rm sub)}  },
\end{aligned}
\end{equation}	
where $\mathcal B(\psi(3686) \to \gamma \chi_{cJ})$ is the branching fraction of $\psi(3686) \to \gamma \chi_{cJ}$, $N_{\psi(3686)}$ is the total number of $\psi(3686)$ events in data, and $\epsilon$ is the detection efficiency. The factor $\mathcal B(\rm sub)$ equals to ${\mathcal B}(\omega \to \pi^{+}\pi^{-}\pi^{0}) \cdot {\mathcal B}(\pi^{0} \to \gamma \gamma)$ for $\chi_{cJ} \to 2K^+ 2K^- \omega$, and to ${\mathcal B}(\phi\to K^+K^-) \cdot {\mathcal B}(\omega \to \pi^{+}\pi^{-}\pi^{0}) \cdot {\mathcal B}(\pi^{0} \to \gamma \gamma)$ for $\chi_{cJ} \to \phi K^{+} K^{-} \omega$~\cite{ref::pdg2025}.
Using this relation, the product branching fractions $\mathcal B(\psi(3686) \to \gamma \chi_{cJ}) \cdot \mathcal B(\chi_{cJ} \to X)$ are determined. The branching fractions $\mathcal B(\chi_{cJ} \to X)$ are obtained by dividing this product by $\mathcal B(\psi(3686) \to \gamma \chi_{cJ})$, taken from the PDG~\cite{ref::pdg2025}. All results for $\chi_{cJ} \to 2K^+ 2K^- \omega$ and $\chi_{cJ} \to \phi K^{+} K^{-} \omega$ are summarized in Table \ref{tab:branching_fraction_phi2k}.

\begin{table*}
    \caption{The signal yields and MC efficiencies for the branching fraction calculations for each signal decay, in which the first uncertainties are statistical and the second systematic.}
\centering
  {
    \begin{tabular}{c|ccc} \hline
      \hline &\multicolumn{3}{c}{$\chi_{cJ} \to 2K^+ 2K^- \omega$}   \\
      \hline             Source               & $\chi_{c 0}$ & $\chi_{c 1}$ & $\chi_{c 2}$    \\
      \hline $N_{\rm net}$     &$1597 \pm 48 $      &$839 \pm 37 $      &$1070 \pm 42 $          \\
       $\epsilon$~(\%)       &$2.64 $        &$3.41$        &$3.41$              \\
       $\mathcal B(\psi(3686) \to \gamma \chi_{cJ}) \cdot \mathcal B(\chi_{cJ} \to 2K^+ 2K^- \omega)$ $(\times 10^{-5})$   &$2.54 \pm 0.08 \pm 0.24$  &$1.03 \pm 0.05 \pm 0.10$  &$1.31 \pm 0.05 \pm 0.13$  \\
       $\mathcal B(\chi_{cJ} \to 2K^+ 2K^- \omega)$ $(\times 10^{-4})$   &$2.60 \pm 0.08 \pm 0.26$  &$1.06 \pm 0.05 \pm 0.11$  &$1.40 \pm 0.06 \pm 0.15$  \\ \hline
      \hline &\multicolumn{3}{c}{$\chi_{cJ} \to \phi K^{+}K^{-} \omega$}   \\ 
      \hline  Source                         & $\chi_{c 0}$ & $\chi_{c 1}$ & $\chi_{c 2}$    \\
      \hline $N_{\rm sig}$     &$949 \pm 37 $      &$501 \pm 28 $      &$724 \pm 33 $          \\
       $N_{\rm sid}$       &$67 \pm 10$        &$36 \pm 9 $       &$71 \pm 11$            \\
       $f$       &$1.05 \pm 0.03$        &$1.03 \pm 0.02 $       &$0.97 \pm 0.02$   \\
       $N_{\rm net}$     &$878 \pm 38 $      &$464 \pm 29 $      &$655 \pm 35 $          \\
       $\epsilon$~(\%)       &$1.71 $        &$2.39$        &$2.49$              \\
       $\mathcal B(\psi(3686) \to \gamma \chi_{cJ}) \cdot \mathcal B(\chi_{cJ} \to \phi K^{+}K^{-} \omega)$ $(\times 10^{-5})$   &$4.39 \pm 0.21 \pm 0.41$  &$1.66 \pm 0.11 \pm 0.16$  &$2.25 \pm 0.13 \pm 0.22$  \\
       $\mathcal B(\chi_{cJ} \to \phi K^{+}K^{-} \omega)$ $(\times 10^{-4})$   &$4.49 \pm 0.20 \pm 0.44$  &$1.70 \pm 0.11 \pm 0.17$  &$2.40 \pm 0.13 \pm 0.24$  \\
      \hline \hline
    \end{tabular}	
    \label{tab:branching_fraction_phi2k}
    }
  \end{table*}

\section{SYSTEMATIC UNCERTAINTY}
\label{sec:systematics}
Systematic uncertainties in the branching-fraction measurements arise from the sources listed in Table~\ref{tab:systematics}. The total uncertainty is obtained by adding all contributions in quadrature.

\begin{table*}[htbp]
	\centering
  	\caption{Relative systematic uncertainties in the measurements of the branching fractions of $\chi_{cJ}\to 2K^+ 2K^- \omega$ and $\chi_{cJ}\to \phi K^{+}K^{-} \omega$ (in \%).}
    \begin{tabular}{c|c|c} \hline
      \hline         & $\chi_{cJ} \to 2K^+ 2K^- \omega$ & $\chi_{cJ} \to \phi K^{+}K^{-} \omega$    \\
      \hline Source    &$\chi_{c0}~/~\chi_{c1}~/~\chi_{c2}$   &$\chi_{c0}~/~\chi_{c1}~/~\chi_{c2}$\\
      \hline $N_{\psi(3686)}$  &0.5         &0.5              \\
       $(K/\pi)^{\pm}$ PID   &6.0         &6.0              \\
       $(K/\pi)^{\pm}$~tracking  &6.0         &6.0              \\
       $\gamma$~selection    &3.0         &3.0              \\
       $\pi^{0}$ reconstruction  &1.0         &1.0              \\
       $\phi$ mass window   &···        &0.2                  \\
       Mass fit        &$0.6~/~1.1~/~1.2$        &$0.2~/~0.4~/~0.8$              \\
       MC generator     &$1.0~/~1.3~/~2.5$        &···                               \\
       Kinematic~fit     &$2.8~/~2.5~/~3.7$        &$2.0~/~3.3~/~2.9$              \\
       MC~statistics    &$0.9~/~0.8~/~0.8$        &$1.1~/~0.9~/~0.9$              \\
       ${\mathcal B}(\psi(3686)\to \gamma\chi_{cJ})$   &2.0~/~2.1~/~2.1      &2.0~/~2.1~/~2.1                     \\
       ${\mathcal B}(\rm sub)$   &0.8          &1.3             \\ \hline
      Total for $\mathcal B(\psi(3686) \to \gamma \chi_{cJ}) \cdot \mathcal B(\chi_{cJ} \to X)$     &$9.7~/~9.6~/~10.1$         &$9.4~/~9.8~/~9.7$              \\
      Total for ${\mathcal B}(\chi_{cJ} \to X)$      &$10.0~/~10.0~/~10.5$         &$9.7~/~10.2~/~10.0$              \\
      \hline \hline
    \end{tabular}
  \label{tab:systematics}
\end{table*}

The total number of $\psi(3686)$ events in data is measured to be $N_{\psi(3686)}=(2712.4 \pm 14.3)\times10^6$ using the inclusive hadronic events, as described in Ref.~\cite{ref::work3}. The uncertainty of $N_{\rm \psi(3686)}$ is 0.5\%.

The systematic uncertainties of the tracking and PID for $(K/\pi)^{\pm}$ have been studied with the control samples of $J/\psi \to K^{*}\bar{K}$ and $J/\psi \to p\bar{p} \pi^{+} \pi^{-}$ in Ref.~\cite{ref::Kpi}. They are assigned as 1.0\% for tracking or PID efficiencies per track.

The photon-detection uncertainty is 1.0\% per photon, and the $\pi^{0}$-reconstruction uncertainty is 1.0\% per $\pi^{0}$, both estimated with $J/\psi\to\pi^+\pi^-\pi^0$~\cite{ref::gamma-recon}.

The systematic uncertainty associated with the $\phi$ mass window is estimated by studying the differences in mass resolution between data and signal MC simulation. The difference of the efficiencies between data and MC simulation is taken as the systematic uncertainty, which is 0.2\%.

The systematic uncertainties of the signal yield extraction are estimated by examining the fits to the invariant mass spectra. 
The uncertainties from the signal shape are estimated by using signal MC simulated shapes convolved with a Gaussian resolution with free parameters. The uncertainties due to the background shape are estimated by varying the order of the Chebyshev polynomial.
The largest differences between the fitted signal yields and the nominal values are taken as the uncertainties. The assigned uncertainties due to the invariant mass fit are 0.6\%, 1.1\%, and 1.2\% for $\chi_{c0,1,2}\to 2K^+ 2K^- \omega$, and 0.2\%, 0.4\%, and 0.8\% for $\chi_{c0,1,2}\to \phi K^{+}K^{-} \omega$, respectively.

The systematic uncertainties due to the mixing of signal MC samples are examined by varying the fractions by $\pm 1 \sigma$. The largest changes in the efficiencies are taken as the systematic uncertainties, which are 1.0\%, 1.3\% and 2.5\% for $\chi_{c0,1,2}\to 2K^+ 2K^- \omega$, respectively.

The systematic uncertainty from the 4C kinematic fit is estimated by comparing the signal efficiencies before and after helix parameter correction, where the correction factors are taken from the previous study~\cite{ref::helixp}.
The changes in the signal efficiencies are assigned as the systematic uncertainties,
which are 2.8\%, 2.5\% and 3.7\% for $\chi_{c0,1,2}\to 2K^+ 2K^- \omega$, and 2.0\%, 3.3\% and 2.9\% for $\chi_{c0,1,2}\to \phi K^{+}K^{-} \omega$, respectively.

The uncertainties due to the limited statistics of the signal MC samples
are 0.9\%, 0.8\%, and 0.8\% for $\chi_{c0,1,2}\to 2K^+ 2K^- \omega$, and 1.1\%, 0.9\%, and 0.9\% for $\chi_{c0,1,2}\to \phi K^{+}K^{-} \omega$, respectively.

The uncertainties from the branching fractions quoted in the PDG~\cite{ref::pdg2025} are 2.0\%, 2.1\%, and 2.1\% for $\psi(3686)\to \gamma\chi_{c0,1,2}$, 0.8\% for ${\mathcal B}(\omega \to \pi^{+}\pi^{-}\pi^{0}) \cdot {\mathcal B}(\pi^{0} \to \gamma \gamma)$, and 1.3\% for ${\mathcal B}(\phi\to K^+K^-) \cdot {\mathcal B}(\omega \to \pi^{+}\pi^{-}\pi^{0}) \cdot {\mathcal B}(\pi^{0} \to \gamma \gamma)$.

\section{Summary}

Using $(2712.4 \pm 14.3)\times10^6$ $\psi(3686)$ events collected with the BESIII detector, the decays $\chi_{c0,1,2}\to 2K^+ 2K^- \omega$ and $\chi_{c0,1,2}\to \phi K^{+}K^{-} \omega$ are observed for the first time with statistical significance exceeding $10\sigma$. The branching fractions are
\begin{gather*}
\mathcal{B}(\chi_{c0}\to 2K^+ 2K^- \omega) = (2.60 \pm 0.08 \pm 0.26) \times 10^{-4}, \\ \\
\mathcal{B}(\chi_{c1}\to 2K^+ 2K^- \omega) = (1.06 \pm 0.05 \pm 0.11) \times 10^{-4}, \\ \\
\mathcal{B}(\chi_{c2}\to 2K^+ 2K^- \omega) = (1.40 \pm 0.06 \pm 0.15) \times 10^{-4},\\ \\ 
\mathcal{B}(\chi_{c0}\to \phi K^{+} K^{-} \omega) = (4.49 \pm 0.20 \pm 0.44) \times 10^{-4}, \\ \\
\mathcal{B}(\chi_{c1}\to \phi K^{+} K^{-} \omega) = (1.70 \pm 0.11 \pm 0.17) \times 10^{-4}, \\ \\
\mathcal{B}(\chi_{c2}\to \phi K^{+} K^{-} \omega) = (2.40 \pm 0.13 \pm 0.24) \times 10^{-4}. 
\end{gather*}
These results expand the known decay modes of $\chi_{cJ}$. Further observations of previously unobserved $\chi_{cJ}$ decays are anticipated, which will improve our understanding of charmonium decay mechanisms.

\section{Acknowledgement}

The BESIII Collaboration thanks the staff of BEPCII (https://cstr.cn/31109.02.BEPC) and the IHEP computing center for their strong support. This work is supported in part by National Key R\&D Program of China under Contracts Nos. 2025YFA1613900, 2023YFA1606000, 2023YFA1606704; National Natural Science Foundation of China (NSFC) under Contracts Nos. 11635010, 11935015, 11935016, 11935018, 12025502, 12035009, 12035013, 12061131003, 12192260, 12192261, 12192262, 12192263, 12192264, 12192265, 12221005, 12225509, 12235017, 12361141819; the Chinese Academy of Sciences (CAS) Large-Scale Scientific Facility Program; the Strategic Priority Research Program of Chinese Academy of Sciences under Contract No. XDA0480600; CAS under Contract No. YSBR-101; 100 Talents Program of CAS; The Institute of Nuclear and Particle Physics (INPAC) and Shanghai Key Laboratory for Particle Physics and Cosmology; ERC under Contract No. 758462; German Research Foundation DFG under Contract No. FOR5327; Istituto Nazionale di Fisica Nucleare, Italy; Knut and Alice Wallenberg Foundation under Contracts Nos. 2021.0174, 2021.0299; Ministry of Development of Turkey under Contract No. DPT2006K-120470; National Research Foundation of Korea under Contract No. NRF-2022R1A2C1092335; National Science and Technology fund of Mongolia; Polish National Science Centre under Contract No. 2024/53/B/ST2/00975; STFC (United Kingdom); Swedish Research Council under Contract No. 2019.04595; U. S. Department of Energy under Contract No. DE-FG02-05ER41374


\end{document}